\documentclass[a4paper]{article}

\usepackage{amsmath}
\usepackage{amssymb}
\usepackage{epsfig}

\newtheorem{teor}{Theorem}
\newtheorem{utv}{Proposition}
\newtheorem{lem}{Lemma}
\newtheorem{sled}{Corollary}

\title{Upper bound on the number of steps for solving the subset sum problem 
by the Branch-and-Bound method}

\author{
        Roman Kolpakov\\
        Lomonosov Moscow State University, Leninskie Gory,\\
				Moscow, 119992 Russia\\
        Email: foroman@mail.ru
				\and
				Mikhail Posypkin\\
        Institute for Information Transmission Problems,\\
        Bolshoy Karetny per., Moscow, 127994 Russia\\
        Email: mposypkin@gmail.com}
\date{}

\begin{document}

\maketitle

\begin{abstract}
We study the computational complexity of one of the particular cases of the knapsack problem: the subset sum problem. 
For solving this problem we consider one of the basic variants of the Branch-and-Bound method in which any sub-problem is 
decomposed along the free variable with the maximal weight. By the complexity of solving a problem by the Branch-and-Bound 
method we mean the number of steps required for solvig the problem by this method. In the paper we obtain upper bounds on 
the complexity of solving the subset sum problem by the Branch-and-Bound method. These bounds can be easily computed from 
the input data of the problem. So these bounds can be used for the the preliminary estimation of the computational resources
required for solving the subset sum problem by the Branch-and-Bound method.
\end{abstract}

\section{Introduction}

The Branch-and-Bound method is one of the most popular approaches to solve global continuous and discrete optimization problems. 
By the complexity of solving a problem by the Branch-and-Bound method we mean the number of decomposition steps (branches) required for solvig the problem
by this method.

In this paper we consider the Branch-and-Bound method for the {\em subset sum problem}. 
The subset sum problem is a particular case of the knapsack problem 
where for each item the price is equal to the weight of the item. 
The subset sum problem  is stated as follows:
\begin{equation}
\label{eq:ssum}
\begin{array}{l}
\mbox{maximize } f(\tilde x) = \sum_{i \in N} x_i w_i,\\
\mbox{subject to } g(\tilde x) = \sum_{i \in N} x_i w_i \leq C,\\
 x_i \in \{0,1\}, i \in N,
\end{array}
\end{equation}
where $N = \{1, \dots, n\}$ is a set of integers between $1$ and $n$, a capacity $C$ and 
weights $w_i$ for $i \in N$ are positive integral numbers.

It is well known that subset sum problem is NP-hard. It means that the worst case complexity 
for the Branch-and-Bound method is an exponential function of $n$. However the number of steps 
may vary significantly for the problems with the same number of variables. 
That is why knapsack algorithms are usually tested on a series of problems generated 
in a different way (see Martello and Toth \cite{Martello} or Kellerer et al. \cite{Kellerer}). 
Knowing the complexity bounds that depend on the problem input coefficients as well the problem 
dimension is very important because such bounds can help to select a proper resolution method 
and estimate resources needed to solve the problem.

Questions of the computational comlexity of boolean programming were actively studied in the literature. 
Jeroslow considered~\cite{Jeroslow} the boolean function maximization problem with equality constraints. 
For the considered problem a wide class of the Branch-and-Bound algorithms was studied, and it was shown 
that the time complexity of solving the problem by any algorithm from this class is $\Omega (2^{n/2})$ 
where $n$ is the number of the problem variables. A similar example of difficult knapsack problem was presented 
in Finkelshein~\cite{Finkel}. It was proved that, for any Branch-and-Bound algorithm solving the considered problem, 
the problem resolution tree contains at least $2 {n + 1 \choose \lfloor n/2 \rfloor + 1} - 1$ nodes where $n$ 
is the number of the problem variables. In Kolpakov and Posypkin~\cite{Kolpak1} the infinite series of knapsack 
instances was constructed which demonstated that the for a particular variant of Branch-and-Bound method proposed 
by Greenberg and Hegerich \cite{Greenberg}, the complexity can be asymptotically $1.5$ times greater 
than $2 {n + 1 \choose \lfloor n/2 \rfloor + 1} - 1$. Thus it was shown that the maximum complexity of solving 
a knapsack problem by the considered method is significantly greater than the lower bound for this value obtained 
in Finkelshtein ~\cite{Finkel}.

The problems proposed by Jeroslow~\cite{Jeroslow} or Finkelshtein~\cite{Finkel} have actually a quite simple form: 
the weights of all the problem variables are equal. Such problems can be easily resolved by the modified 
Branch-and-Bound method enchanced with the the dominance relation. Paper Chvatal~\cite{Chvatal} dealt
with recursive algorithms that use the dominance relation and improved linear relaxation to reduce the enumeration. 
The author suggested a broad series of problems unsolvable by such algorithms in a polynomial time.

The Branch-and-Bound complexity for integer knapsack problems were considered by Aardal~\cite{Aardal} and  
by Krishnamoorthy~\cite{Krishna}.  Several papers were devoted to obtaining upper bounds on complexity of
solving boolean knapsack problems by the Branch-and-Bound method. In Grishuknin~\cite{Grish}  
an upper bound on the complexity of solving a boolean knapsack problem by the majoritarian Branch-and-Bound 
algorithm was proposed. This bound depends only on the number of problem variables $n$ and 
ignores problem coefficients. In Kolpakov and Posypkin~\cite{Kolpak2} upper bounds for the complexity 
of solving a boolean knapsack problem by the Branch-and-Bound algorithms with an arbitrary choice 
of decomposition variable were obtained. Unlike bounds proposed in Girshukhin~\cite{Grish}, these bounds 
take into account both problem size and coefficients. 


\section{Preliminaries}
\label{sec:pre}

A boolean tuple $\tilde x=(x_1, x_2,\ldots , x_n)$ such that $g(\tilde x) \leq C$
is called a {\em feasible solution} of the problem~(\ref{eq:ssum}).
A feasible solution~$\tilde x$ of a problem (\ref{eq:ssum}) is called an {\em optimal solution} 
if for any  other feasible solution $\tilde y$ of the problem~(\ref{eq:ssum}) 
the inequality $f(\tilde y) \leq f(\tilde x)$ holds.
Solving the problem~(\ref{eq:ssum}) means finding at least one of its optimal solutions.

We define a {\em map} as a pair $(I, \theta)$ of a set $I\subseteq N$ and a mapping $\theta: I \to \{0, 1\}$. 
Any map $(I, \theta)$ defines a {\em subproblem} formulated as follows:
\begin{equation}
\label{eq:sub}
\begin{array}{l}
\mbox{maximize } f(x) = \sum_{i \in N} w_i x_i,\\
\mbox{subject to } g(x) = \sum_{i \in N} w_i x_i \leq C,\\
x_i = \theta(i), i \in I, \\
x_i \in \{0, 1\}, i \in N \setminus I.
\end{array}
\end{equation}
The set $\{ x_i\;:\; i\in I\}$ is called the {\em set of fixed variables} of the subproblem~(\ref{eq:sub}). 
The set $\{ x_i\;:\; i\in N \setminus I\}$ is called the {\em set of free variables} of this subproblem.

In the sequel we will refer to the subproblem~(\ref{eq:sub}) as the {\em respective} or {\em corresponding} subproblem 
for the map $(I, \theta)$ and 
will refer to the map $(I, \theta)$ as  the {\em respective} or {\em corresponding} map for  subproblem~(\ref{eq:sub}).

A boolean tuple $\tilde x=(x_1, x_2,\ldots , x_n)$ such that
\[
\begin{array}{l}
g(x) \leq C,\\
x_i = \theta(i), i \in I,
\end{array}
\]
is called {\em a feasible solution} of subproblem~(\ref{eq:sub}).
Clearly, any {\em feasible solution} of subproblem~(\ref{eq:sub}) 
is a feasible solution of problem~(\ref{eq:ssum}) as well.
A feasible solution~$\tilde x$ of subproblem (\ref{eq:sub}) is called {\em optimal} 
if for any  other feasible solution~$\tilde y$ of this
subproblem the inequality $f(\tilde y) \leq f(\tilde x)$ holds.

For any map $z = (I, \theta)$ define its {\em $1$-complement} $\tilde{z}^{(1)}$ 
as a tuple $(z^{(1)}_1, z^{(1)}_2,\ldots , z^{(1)}_n)$ such that
\[
z^{(1)}_i = \begin{cases} \theta(i), i \in I,\\ 1, i \in N \setminus I. \end{cases}
\]

The {\em $0$-complement} $\tilde{z}^{(0)}=(z^{(0)}_1, z^{(0)}_2,\ldots , z^{(0)}_n)$ of the map~$z$ is defined as follows:
\[
z^{(0)}_i = \begin{cases} \theta(i), i \in I,\\ 0, i \in N \setminus I. \end{cases}
\]

Let $W = \sum_{i \in N} w_i$.
subproblem~(\ref{eq:sub}) satisfies {\em C0-condition} if $\sum_{i \in I} \theta(i) w_i > C$ and 
satisifies {\em C1-condition} if $\sum_{i \in I} (1 - \theta(i)) w_i \geq W - C$.
This following statement is an immediate consequence of the C0-condition definition.

\begin{utv}
\label{utv:c0}
A subproblem~(\ref{eq:sub}), satisfying C0-condition, has no feasible solutions.
\end{utv}

\begin{utv}
\label{utv:c1}
If a subproblem~(\ref{eq:sub}) satisfies C1-condition then the $1$-complement of the respective map $(I, \theta)$ is 
an optimal solution for this subproblem.
\end{utv}

{\bf Proof.} Let a subproblem~(\ref{eq:sub}) satisfy C1-condition, and $\tilde{z}^{(1)}$ be the $1$-complement 
of the map $z=(I, \theta)$. 
Then
\[
 C \geq W - \sum_{i \in I} (1 - \theta(i)) w_i =
\sum_{i \in I} \theta(i) w_i + \sum_{i \in N\setminus I} w_i = \sum_{i \in N} z^{(1)}_i w_i.
\]
Therefore $\sum_{i \in N} z^{(1)}_i w_i \leq C$, so $\tilde{z}^{(1)}$ is a feasible solution of the subproblem~(\ref{eq:sub}). 
Since in $\tilde{z}^{(1)}$ all variables from the set $N \setminus I$ take the value~$1$, the solution $\tilde{z}^{(1)}$ is 
obviously optimal.$\Box$

This corollary immediately follows from Propositions \ref{utv:c0} and~\ref{utv:c1}.
\begin{sled}
\label{sled:c01}
A subproblem (\ref{eq:sub}) can not satisfy both C0-condition and C1-condition at the same time.
\end{sled}

\begin{utv}
\label{utv:c0c1}
If $I = N$ then subproblem~(\ref{eq:sub}) satisfies either C0-condition or C1-condition.
\end{utv}
{\bf Proof.}
Consider subproblem (\ref{eq:sub}) such that $I = N$. Assume that subproblem (\ref{eq:sub}) does not satisfy C0-condition: 
$\sum_{i\in N} \theta(i) w_i \leq C$. Since $\sum_{i\in N} \theta(i) w_i = W - \sum_{i\in N} (1 - \theta(i))w_i$, in this case
we have $W - \sum_{i\in N} (1 - \theta(i)) w_i \leq C$, i.e. subproblem~(\ref{eq:sub}) satisfies C1-condition.$\Box$

For a map $z = (I, \theta)$, where $I \neq N$ and $i \in N \setminus I$, we introduce two new maps 
$z_0 = (I', \theta_0)$, $z_1 = (I', \theta_1)$  where  $I' = I \cup i$  and
\[
\theta_k(j) = \begin{cases} \theta(j) \mbox{ for } j \in I, \\ k \mbox{ for } j = i,\end{cases}
k = 0,1.
\]
The set of the two subproblems corresponding to the maps $z_0$ and~$z_1$ is called 
the {\em decomposition of subproblem~(\ref{eq:sub}) along the variable~$x_i$}. 
For this decomposition the variable $x_i$ is called the {\em split variable} and 
the index $i$ is called the {\em split index}.

\begin{utv}
\label{utv:decomp}
Let $\{P_0, P_1\}$ be a decomposition of a subproblem~$P$ along some variable.
Then the set of feasible (optimal) solutions of the subproblem~$P$ is
the union of the sets of feasible (optimal) solutions of the subproblems $P_0$ and~$P_1$.
\end{utv}

\section{The Branch-and-Bound Algorithm}
\label{sec:bnb}

In this paper we study one of the basic variants of the Branch-and-Bound algorithm
for solving the subset sum problem which we call the {\em majoritarian} Branch-and-Bound (MBnB) 
algorithm.

{\bf\underline{MBnB algorithm}}

During the execution the algorithm maintains the list~${\cal S}$ of subproblems waiting for processing
and the {\em incumbent solution} $\tilde x_r$. The incumbent solution is the best feasible solution found so far.

{\bf Step 1.} The list~${\cal S}$ of subproblems is initialized by the original problem~(\ref{eq:ssum}): 
${\cal S} = \{ P_0\}$, where $P_0$ is the original problem. All components of the incumbent solution are set to zero.

{\bf Step 2.} An arbitrary subproblem $P$ in the list ${\cal S}$ is selected for processing and is removed from this list.

{\bf Step 3.} Three cases for processing~$P$ are possible:
\begin{itemize}

\item The subproblem $P$ satisfies C0-condition. Then, by Proposition~\ref{utv:c0}, 
$P$ does not have feasible solutions and thus can be safely excluded from the further processing.

\item The sub-roblem $P$ satisfies C1-condition. In that case the $1$-complement $\tilde{z}^{(1)}$ 
of the map $z$ corresponding to the subproblem~$P$ is compared with the incumbent solution
(recall that, by Proposition~\ref{utv:c0}, $\tilde{z}^{(1)}$ is an optimal solution for~$P$). 
If $f(\tilde{z}^{(1)}) > f(\tilde x_r)$ then the incumbent solution is replaced by $\tilde{z}^{(1)}$.

\item The subproblem $P$ satisfies neither C0-condition nor C1-condition. Then the subproblem~$P$ is decomposed 
along the variable $x_i$ where $x_i$ is the free variable of~$P$ with the maximal weight~$w_i$, i.e.
$i = \mbox{argmax}_{j \in N \setminus I} w_j$. The two subproblems of the decomposition are added to the list~${\cal S}$.
\end{itemize}

{\bf Step 4.} If the list ${\cal S}$ is empty the algorithm terminates. Otherwise the algorithm continues from the step 2.

Since the number of variables of the original problem is finite the MBnB algorithm terminates in a finite number of steps. 
It follows from Propositions \ref{utv:c0}-\ref{utv:decomp}  that the resulting incumbent solution is an optimal solution of 
the original problem.

Note that in the MBnB algorithm any subproblem is decomposed along the free variable with the maximal weight.
So without loss of generality we will assume that all variables $x_1, x_2,\ldots , x_n$ of the original 
problem~(\ref{eq:ssum}) are ordered in the non-increasing order of their weights, i.e. $w_1\ge w_2\ge\ldots\ge w_n$.
In this case any subproblem is decomposed along the free variable with the minimal index, 
i.e. for any decomposed subproblem~(\ref{eq:ssum}) we have $I=\{1, 2,\ldots ,s\}$ where $0\le s<n$,
and $x_{s+1}$ is the split variable for the subproblem decomposition.

The problem resolution process can be represented as a directed {\em MBnB-tree}. 
The subproblems processed by the MBnB algorithm form the set of tree nodes. 
Each subproblem decomposed by the MBnB algorithm is connected by directed arcs
with the two subproblems constituting its decomposition.  
The {\em root} of the MBnB-tree corresponds to the original problem~(\ref{eq:ssum}). Obviously the number of iterations of the main loop of the MBnB algorithm equals to the number of nodes in the respective MBnB tree. 
The {\em MBnB complexity of the problem~(\ref{eq:ssum})} is defined as the number of iterations of the main loop of the MBnB algorithm required to resolve the problem (the total number of nodes in the MBnB tree). 
Notice that the total number of nodes in the MBnB tree can be computed as $2 L - 1$, where $L$ is a number of leaf nodes in the MBnB tree.

Leaf nodes of the MBnB-tree correspond to subproblems satisfying either C0-condition or C1-condition. 
The leaf nodes are marked by tuples as follows:
\begin{itemize}
\item a leaf node corresponding to a subproblem satisfying C0-condition is marked by the $0$-complement 
of the map corresponding to the subproblem, such tuples are called {\em leaf $0$-tuples};
\item a leaf node corresponding to a subproblem satisfying C1-condition is marked by the $1$-complement 
of the map corresponding to the subproblem, such tuples are called {\em leaf $1$-tuples}
\end{itemize}

\begin{figure}
\begin{center}
\epsfig{file=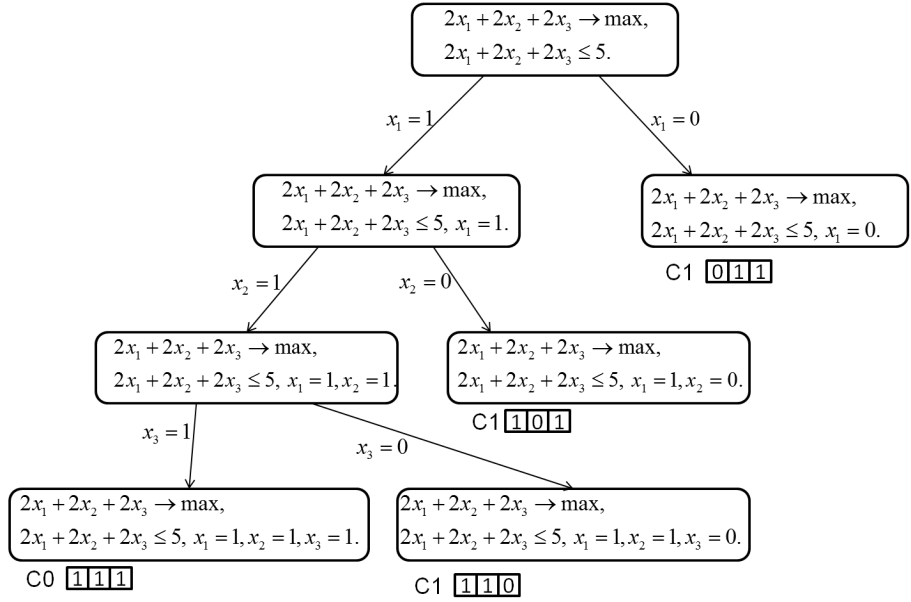, width=12cm}
\caption{Sample MBnB tree}
\label{fig:tree}
\end{center}
\end{figure}

As an example, the MBnB-tree for the subset sum problem
\[
f(x) = 2x_1 + 2x_2 + 2x_3 \to max,
2x_2 + 2x_2 + 2x_3 \leq 5,
\]
is depicted at Figure~\ref{fig:tree}. Each leaf node is marked with the respective C0- or C1-condition 
and the assigned $0$- or $1$-tuple. 
The MBnB  complexity of this problem is seven.

Consider the set~$B^n$ of binary tuples of a length $n$ over the set $\{0, 1\}$. 
Define the partial order in~$B^n$ as follows: 
$\tilde\alpha \leq \tilde\beta$ if $\alpha_i \leq \beta_i$ for all $i \in N$. 
If $\tilde\alpha \leq \tilde\beta$ does not hold we write $\tilde\alpha \not \leq \tilde\beta$.

\begin{utv}
All leaf $0$-tuples  are pairwise incomparable.
\label{0antichain}
\end{utv}
{\bf Proof.} Let $\tilde\alpha=(\alpha_1,\ldots ,\alpha_n)$ and $\tilde\beta=(\beta_1,\ldots ,\beta_n)$ 
be two different $0$-tuples such that $\tilde\alpha \leq \tilde\beta$,
and $P_\alpha$ ($P_\beta$) be  the subproblem respective for $\tilde\alpha$ ($\tilde\beta$). 
There exists $j \in N$ such that $\alpha_j = 0$ and $\beta_j = 1$. 
According to the leaf $0$-tuple definition, $\sum_{i \in N} \alpha_i w_i > C$. Thus
\begin{equation}
\label{eq:eq1}
\sum_{i \in N} \beta_i w_i \geq \sum_{i \in N} \alpha_i w_i + w_j > C + w_j.
\end{equation}
Note that $x_j$ is a fixed variable of $P_\beta$ because all free variables of $P_\beta$ have 
zero values in $\tilde\beta$.

Let $P_\beta$ be resulted from the decomposition of some subproblem $P$ along a variable $x_k$. 
Since the decomposition is always performed along the free variable with the maximal weight, we
have that $w_k = \min_{x_i\in F_\beta} w_i$, where $F_\beta$ is the set of fixed variables of 
the subproblem $P_\beta$. Thus $w_k \leq w_j$ because $x_j \in F_\beta$. The subproblem $P$ 
does not correspond to a leaf node and thus it does not satisfy C0-condition, i.e. 
$\sum_{x_i \in F_\beta \setminus \{x_k\}} \beta_i w_i \le C$. Therefore
\begin{equation}
\label{eq:eq2}
\sum_{i \in N} \beta_i w_i = \sum_{x_i \in F_\beta} \beta_i w_i \le C + w_k \leq C + w_j.
\end{equation}

Inequalities (\ref{eq:eq1}) and (\ref{eq:eq2}) contradict each other. 
Thus the proposition is proved.$\Box$

In the same way we can prove the following statement.

\begin{utv}
All leaf $1$-tuples are pairwise incomparable.
\label{1antichain}
\end{utv}

Following to Propositions \ref{0antichain} and~\ref{1antichain}, the set of all leaf $0$-tuples 
is called the {\em $0$-antichain} and the set of all leaf $1$-tuples is called the {\em $1$-antichain}.

\begin{utv}
\label{utv:01chains}
If $\tilde\alpha$ is a leaf $0$-tuple and $\tilde\beta$ is a leaf $1$-tuple 
then $\tilde\alpha \not\leq \tilde\beta$.
\end{utv}
{\bf Proof.} Since, by Proposition~\ref{utv:c1}, $\tilde\beta$ is a feasible solution for the subproblem 
marked by~$\tilde\beta$, the inequality $f(\tilde\beta) \leq C$ holds. 
By the definition of leaf $0$-tuple, the subproblem marked by~$\tilde\alpha$ satisfies C0-condition.
Hence, by Proposition~\ref{utv:c0}, $\tilde\alpha$ is not a feasible solution for this subproblem,
i.e. $f(\tilde\alpha) > C$. Thus $f(\tilde\beta) < f(\tilde\alpha)$. 
Therefore $\tilde\alpha \not\leq \tilde\beta$ because the function $f$ is obviously non-decreasing
 w.r.t. the introduced order in~$B^n$.$\Box$

\section{Basic properties of binary tuples}
\label{sec:binary}

This section entirely focuses on the binary tuples and their properties. 
The obtained results will be used at the end of the paper for finding 
the upper bound for the MBnB complexity of the subset sum problem.

\subsection{Connected components}

Let $\tilde\alpha=(\alpha_1,\ldots,\alpha_n)$ be a binary tuple from $B^n$.
We call a component $\alpha_i$ of $\tilde\alpha$ {\em 1-component} 
({\em 0-component}) if $\alpha_i=1$ ($\alpha_i=0$). 
The number of 1-components in $\tilde\alpha$ is called {\em the weight} 
of $\tilde\alpha$ and is denoted by $\|\tilde\alpha\|$.

We denote by $B_+^n$ the set of all binary tuples from $B^n$ in which the number 
of 1-components is greater than the number of 0-components, 
i.e. $B_+^n = \{\tilde\alpha \in B^n : \|\tilde\alpha\| > n/2\}$.

For $1\leq i,j\leq n$ we denote by $\tilde\alpha [i:j]$ the tuple $(\alpha_i,\ldots,\alpha_j)$ 
if $i\le j$ and the tuple $(\alpha_1,\ldots,\alpha_j,\alpha_i,\ldots,\alpha_n)$ if $i>j$.
Such tuples are called {\em segments}. 

The segment $\tilde\alpha [i:j]$ {\em precedes}  the component $\alpha_{j+1}$ ($\alpha_1$) if $j<n$ ($j=n$). 
The segment $\tilde\alpha [i:j]$ {\em succeeds}  the component $\alpha_{i-1}$ ($\alpha_n$) if $i>0$ ($i=0$). 
If $i < j$, a {\em prefix (suffix) of the segment} $\tilde\alpha [i:j]$ is any segment $\tilde\alpha [i:j']$ 
($\tilde\alpha [i':j]$) where $i\le j' < j$ ($i < i'\le j$). If $i > j$ a {\em prefix (suffix) of the segment} 
$\tilde\alpha [i:j]$ is any segment $\tilde\alpha [i:j']$ ($\tilde\alpha [i':j]$) where  $i\le j'\le n$ 
($i < i' \le n$) or $1\le j' < j$ ($1 \le i' \le j$).

A segment is called {\em balanced} if in this segment the number of 0-components is equal to the number of 1-components. 
A segment is called {\em 0-dominated} ({\em 1-dominated}) if in this segment the number of 0-components 
is greater (is less) than the number of 1-components. 
A balanced segment is called {\em a minimal balanced segment} if any prefix of this segment is 0-dominated. 
There is obviously the equivalent definition: a balanced segment is called {\em minimal balanced segment} 
if any suffix of this segment is 1-dominated.

A 0-component $\alpha_i$ is called {\em connected} to a 1-component $\alpha_j$ and a 1-component $\alpha_j$ 
is called {\em connected} to a 0-component $\alpha_i$ if the segment $\tilde\alpha[i, j]$ is a minimal balanced segment. 

\begin{figure}
\begin{center}
\epsfig{file=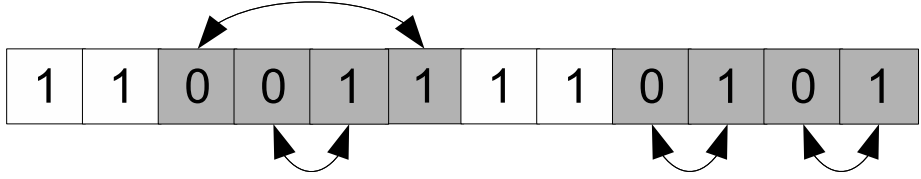, width=12cm}
\caption{Bound components of a tuple}
\label{fig:bound}
\end{center}
\end{figure}

All components of a tuple  connected to some other components are called {\em bound}. 
All other components are called {\em unbound}. At Figure~\ref{fig:bound} bound components 
are shadowed and the connection between components is visualized by arcs. 

The following statement is almost obvious:

\begin{utv}
\label{utv:connrel}
\begin{enumerate}
\item Any bound 0-component is connected to exactly one 1-component and similarly,
any bound 1-component is connected to exactly one 0-component.
\item The set of all bound components is a union of pairs consisting of one 0-component 
and one 1-component connected to each other.

\end{enumerate}
\end{utv}

\begin{utv}
\label{utv:conn}
If a 0-component $\alpha_i$ is connected to a 1-component $\alpha_j$ 
then any component in the segment $\tilde\alpha[i:j]$ 
is connected to another component in the same segment.
\end{utv}
{\bf Proof.} Let $\alpha_k$ be a 1-component from $\tilde\alpha[i:j]$, $k \neq j$. 
By the definition of connected components the segment $\tilde\alpha[i:j]$ is a minimal 
balanced segment and hence the segment $\tilde\alpha[i:k]$ is 0-dominated. 
But $\tilde\alpha[k:k]$ is a 1-dominated segment and thus there should exist at least one balanced 
suffix of the segment $\tilde\alpha[i:k]$. Choose in $\tilde\alpha[i:k]$ the balanced suffix $\tilde\alpha[l:k]$ 
of the minimal length. Clearly, any suffix of the segment $\tilde\alpha[l:k]$ is 1-dominated. 
So $\tilde\alpha[l:k]$ is a minimal balanced segment. Thus $\alpha_l$ is a 0-component 
connected to the 1-component $\alpha_k$. 
In a similar way it can be proved that any 0-component from the segment $\tilde\alpha[i:j]$ is connected 
to some 1-component in the same segment.$\Box$

A segment $\tilde\alpha[i:j]$ is called {\em a connected segment} 
if $\alpha_i$ and $\alpha_j$ are connected to each other. 
A connected segment $\tilde\alpha[i:j]$ is called {\em maximal} 
if there is no other connected segment containing $\tilde\alpha[i:j]$. 
The following result is then obvious.

\begin{utv}
\label{utv:consegbal}
Any connected segment is balanced.
\end{utv}

Two components of a tuple are {\em neighbouring} if their indexes differ by one. 
Moreover, the first and the last components of a tuple are also assumed to be neighbouring. 
Two segments of a tuple are {\em separated} if one of these segments has no components 
neighbouring with components of the other segment.
As any other subset of components in a tuple, the set of all bound components in $\tilde\alpha$ is 
a union of pairwise separated segments. We call these segments {\em bound segments} of $\tilde\alpha$. 
The tuple depicted at Figure~\ref{fig:bound} has two bound segments. 
From Proposition~\ref{utv:conn} we conclude

\begin{utv}
\label{utv:bound1}
Any bound segment is a union of one or more  non-overlapping maximal connected segments.
\end{utv}

Using this statement and Proposition~\ref{utv:consegbal}, 
it is not difficult to prove the following fact.

\begin{utv}
Any bound segment is balanced, and  any prefix (suffix) of any bound segment is 0-dominated (1-dominated) or balanced.
\label{onpref}
\end{utv}

The following criterion takes place.
\begin{lem}
A 1-component is bound in~$\tilde\alpha$ if and only if there
exists a 0-dominated segment  preceding this component  in~$\tilde\alpha$.
\label{on1comp}
\end{lem}
{\bf Proof.}
Let $\alpha_i$ be a bound 1-component in~$\tilde\alpha$.  Then there should exist 
a minimal balanced segment $\tilde\alpha[i':i]$ such that $i'<i$ and $\alpha_i' = 0$. 
Clearly, the segment $\tilde\alpha[i',i-1]$ precedes $\alpha_i$ and is 0-dominated. 
Thus the necessity is proved. To prove the sufficiency, assume
that there exists some 0-dominated segment $\tilde\alpha[i':i-1]$ preceeding $\alpha_i$. 
Then the number of 0-components in the segment $\tilde\alpha[i':i]$ is not less than 
the number of 1-components in this segment. Let the segment $\tilde\alpha[i':i]$ be
also 0-dominated. Then, since the segment $\tilde\alpha[i:i]$ is 1-dominated, 
there should exist at least one balanced suffix in the segment $\tilde\alpha[i':i]$. 
Thus, in any case there exists at least one balanced segment $\tilde\alpha[l:i]$
such that $i'\le l<i$. Let $\tilde\alpha[i'', i]$ be the such segment of the minimal length. 
Obviously, $\tilde\alpha[i'', i]$ is a minimal balanced segment. So $\alpha_i$ is a bound 
component connected to $\alpha_i''$.$\Box$

The following corollary is a direct consequence of the Lemma~\ref{on1comp}.
\begin{sled}
Let $\tilde\alpha=(\alpha_1,\ldots,\alpha_n)$, $\tilde\alpha'=(\alpha'_1,\ldots,\alpha'_n)$
be two tuples from $B_+^n$ such that $\tilde\alpha\le\tilde\alpha'$, and $\alpha_i$ be
a 1-component unbound in~$\tilde\alpha$. Then $\alpha'_i$ is a 1-component unbound in~$\tilde\alpha'$.
\label{sloncomp1}
\end{sled}

Similarly to Lemma~\ref{on1comp} one can prove the following lemma:

\begin{lem}
A 0-component is bound in~$\tilde\alpha$ if and only if in~$\tilde\alpha$ there
exists a 1-dominated segment  succeeding this component.
\label{on0comp}
\end{lem}

From this lemma we easily obtain the following corollary.

\begin{sled}
All 0-components in any tuple from $B^n_+$ are bound.
\label{bnplus}
\end{sled}

\subsection{Projection mapping ${\cal D}$}

Define the {\em projection mapping} ${\cal D}: B_+^n \to B^n$ as follows.
Let $\tilde\alpha\in B_+^n$. Among all unbound 1-components  in $\tilde\alpha$ 
choose the component with the maximal index. 
We denote by ${\cal D}(\tilde\alpha)$ the binary tuple obtained from $\tilde\alpha$ 
by replacing this component with~$0$. 

We have immediately from Corollary~\ref{sloncomp1} 
\begin{sled}
Let $\tilde\alpha=(\alpha_1,\ldots,\alpha_n)$, $\tilde\alpha'=(\alpha'_1,\ldots,\alpha'_n)$
be two tuples from $B_+^n$ such that $\tilde\alpha = {\cal D}(\tilde\alpha')$, and
$\alpha_i$ be an unbound 1-component  in $\tilde\alpha$. Then $\alpha'_i$ is
an unbound 1-component in $\tilde\alpha'$.
\label{sloncomp2}
\end{sled}

The following lemma states that the operation ${\cal D}$ is injective.
\begin{lem}
For any two different tuples $\tilde\alpha',\tilde\alpha''$ from $B_+^n$  the tuples 
${\cal D}(\tilde\alpha')$, ${\cal D}(\tilde\alpha'')$ are also different.
\label{injectD}
\end{lem}
{\bf Proof.}
 Let  $\tilde\alpha'=(\alpha'_1,\ldots,\alpha'_n)$, 
$\tilde\alpha''=(\alpha''_1,\ldots,\alpha''_n)$ be two arbitrary different tuples from $B_+^n$.  
Let ${\cal D}(\tilde\alpha')$ 
be obtained from $\tilde\alpha'$ by substitution of zero for a 1-component
$\alpha'_{i'}$, and ${\cal D}(\tilde\alpha'')$ be obtained from $\tilde\alpha''$
by substitution of zero for a 1-component $\alpha''_{i''}$. If $i' = i''$ then 
${\cal D}(\tilde\alpha') \neq {\cal D}(\tilde\alpha'')$ and so the lemma is valid.  

Consider the case  $i'\neq i''$. Without loss of generality assume that $i'' < i'$.
To prove this case, assume also that ${\cal D}(\tilde\alpha')={\cal D}(\tilde\alpha'')$. 
Then $\alpha'_{i''} = 0$. Therefore, if $i' = i'' + 1$ then $\alpha'_{i'}$ is obviously 
connected to $\alpha'_{i''}$, i.e. $\alpha'_{i'}$ is bound in $\tilde \alpha'$.
Thus $i' > i'' + 1$. 

Since $\alpha''_{i''}$ has the maximal index 
among all components unbound in $\tilde\alpha''$, the components
$\alpha''_{i''+1},\ldots,\alpha''_{i'-1}$ are bound in $\tilde\alpha''$,
i.e. the segment $\tilde\alpha'' [i''+1:i'-1]$ is a 
prefix of some bound segment of $\tilde\alpha''$. Therefore, 
by Proposition~\ref{onpref}, in $\tilde\alpha'' [i''+1:i'-1]$ 
the number of 1-components is not greater than the number of 0-components. 
Since ${\cal D}(\tilde\alpha')={\cal D}(\tilde\alpha'')$ 
all components of the tuple $\tilde\alpha'$ except $\alpha'_{i'}$ and $\alpha'_{i''}$ 
have to coincide with the respective components of the tuple $\tilde\alpha''$.
Hence the segment $\tilde\alpha' [i''+1:i'-1]$ has to coincide with the
segment $\tilde\alpha'' [i''+1:i'-1]$. So in $\tilde\alpha' [i''+1:i'-1]$
the number of 1-components is not also greater than the number of 0-components.
Moreover, it follows from ${\cal D}(\tilde\alpha')={\cal D}(\tilde\alpha'')$
that $\alpha'_{i''}=0$. Thus, in the segment $\tilde\alpha' [i'':i'-1]$
of $\tilde\alpha'$ the number of 0-components is greater than the number of 
1-components. Therefore, by Lemma~\ref{on1comp} the component $\alpha'_{i'}$
is bound in $\tilde\alpha'$. That contradicts our assumption that $\alpha'_{i'}$
is unbound in $\tilde\alpha'$.

For $s > n/2$ define the tuple $\tilde\gamma_s = (\underbrace{0,\ldots,0}_{n-s},
\underbrace{1,\ldots,1}_s)$.

\begin{lem}
If for a tuple $\tilde\alpha\in B_+^n$, where $\|\tilde\alpha\|\ge n/2+1$, the
relation $\tilde\alpha\not\le\tilde\gamma_s$ is valid, then
${\cal D}(\tilde\alpha)\not\le\tilde\gamma_s$.
\label{nesravD1}
\end{lem}
{\bf Proof.}
Let $\tilde\alpha=(\alpha_1,\ldots,\alpha_n)$. Consider separately two cases: 
$\alpha_1=0$ and $\alpha_1=1$. Let $\alpha_1=0$. 
Let $\alpha_i$ be the 1-component with the minimal index in $\tilde\alpha$.
Note that $i>1$ and $\alpha_{i-1} = 0$. It is obvious that the $0$-component 
$\alpha_{i-1}$ is connected to the $1$-component $\alpha_i$. 
Thus $\alpha_i$ is bound in~$\tilde\alpha$. According to the definition of ${\cal D}$, 
$\alpha_i$ coincides with the respective component of ${\cal D}(\tilde\alpha)$,
i.e. the $i$-th component of ${\cal D}(\tilde\alpha)$ is an 1-component.
Note that $i\le n-s$, since otherwise $\tilde\alpha\le\tilde\gamma_s$.
Thus, we obtain that ${\cal D}(\tilde\alpha)\not\le\tilde\gamma_s$. 

Now let $\alpha_1=1$. From $\tilde\alpha\not\le\tilde\gamma_s$ 
we have $s<n$. If $\alpha_1$ is bound in~$\tilde\alpha$ then 
$\alpha_i$ coincides with the respective first component of ${\cal D}(\tilde\alpha)$,
so the first component of ${\cal D}(\tilde\alpha)$ is an 1-component. 
Therefore ${\cal D}(\tilde\alpha)\not\le\tilde\gamma_s$ in this case. 
Let $\alpha_1$ be unbound in $\tilde\alpha$. Note that the condition 
$\|\tilde\alpha\|> n/2+1$ implies that in~$\tilde\alpha$ at least two 
components are unbound. So $\alpha_1$ cannot be the component with 
the maximal index among all components unbound in~$\tilde\alpha$. Hence, 
by the definition of the tuple ${\cal D}(\tilde\alpha)$, its first component 
has to coincide with $\alpha_1$. Thus, this component has to be an 1-component 
which implies ${\cal D}(\tilde\alpha)\not\le\tilde\gamma_s$.  $\Box$

\begin{lem}
Let for a tuple $\tilde\alpha\in B_+^n$ the relations 
$\tilde\alpha\not\le\tilde\gamma_s$ 
and ${\cal D}(\tilde\alpha)\le\tilde\gamma_s$ be valid. Then there is no such tuple 
$\tilde\alpha' \in B^n$ that $\tilde\alpha = {\cal D}(\tilde\alpha')$.
\label{nesravD2}
\end{lem}
{\bf Proof.}
Assume that $\tilde\alpha=(\alpha_1,\ldots,\alpha_n)$, and the relations $\tilde\alpha\not\le\tilde\gamma_s$ 
and ${\cal D}(\tilde\alpha)\le\tilde\gamma_s$ are valid. Then it follows from Lemma~\ref{nesravD1} 
that $n$ is odd and $\|\tilde\alpha\| =(n+1)/2$. Thus there is only one unbound 1-component in~$\tilde\alpha$. 
Let $\alpha_i$ be this component. Since ${\cal D}(\tilde\alpha)\le\tilde\gamma_s$ and 
$\tilde\alpha\not\le\tilde\gamma_s$, the component $\alpha_i$ has to be the only 1-component 
in~$\tilde\alpha$ satisfying the condition $i\le n-s$. Therefore, if $i>1$ then $\alpha_i$ is
obviously connected to the 0-component $\alpha_{i - 1}$, which contradicts our assumption that 
$\alpha_i$ is unbound. Consider the only possible case $i=1$.

Assume that there exists a tuple $\tilde\alpha'=(\alpha'_1,\ldots,\alpha'_n)$  
such that $\tilde\alpha = {\cal D}(\tilde\alpha')$. Since $\alpha_1$ is unbound 
in~$\tilde\alpha$, by Corollary~\ref{sloncomp2} the component $\alpha'_1$ is unbound 
in $\tilde\alpha'$. Let $\alpha'_j$ be the 1-component of $\tilde\alpha'$ replaced 
by zero in~$\tilde\alpha$. Clearly, $j \neq 1$. According to the definition of 
the mapping ${\cal D}$, $\alpha'_j$ is the unbound component with the maximal index
in $\tilde\alpha'$. Since $\alpha'_1$ is unbound in $\tilde\alpha'$, we have two possible cases: 
$j=n$ or $\tilde\alpha' [j+1:n]$ is a bound segment of~$\tilde\alpha'$.
In the first case $\alpha_1$ should be obviously connected to the 0-component $\alpha_0$,
i.e. $\alpha_1$ is bound in $\tilde\alpha$. This contradicts our assumption. 
In the second case by Proposition~\ref{onpref} the segment $\tilde\alpha'[j+1:n]$ 
is balanced. Therefore, since the segments $\tilde\alpha'[j+1:n]$ and $\tilde\alpha[j+1:n]$ 
are identical and $\alpha_j = 0$, the segment $\tilde\alpha[j:n]$ is 0-dominated. 
Hence, by Lemma~\ref{on1comp}, the component $\alpha_1$ has to be bound in~$\tilde\alpha$ 
which contradicts again our assumption. $\Box$

\subsection{Properties of antichains in~$B^n$}

Let $T', T''$ be two antichains in $B^n$ such that $\tilde\alpha'\not\ge\tilde\alpha''$
for any tuples $\tilde\alpha'\in T'$, $\tilde\alpha''\in T''$. We will denote
this case by $T'<T''$ (note that $T'<T''$ implies $T'\cap T''=\emptyset$). 
For $s>n/2$ denote by ${\cal A}_s$ the set of all pairs $(T', T'')$ of antichains 
in $B^n$ such that $T'<T''$ and $\tilde\gamma_s\in T'$. The cardinality of a pair 
of non-overlapping antichains is the total number of tuples in these antichains.

Recall that the number of 1-components in a binary tuple~$\tilde\alpha$ from $B^n$ is called 
the {\em weight} of this tuple and is denoted by $\|\tilde\alpha\|$. By $B^n_k$ we denote the 
set of all tuples from $B^n$ whose weights are equal to~$k$. A pair of antichains $(T', T'')$ 
from ${\cal A}_s$ is called {\em regular from below} if in the set $\left(T'\cup T''\right)\setminus\{\tilde\gamma_s\}$ 
all tuples of minimum weight are contained in $T'$ and is called {\em regular from above}
if in the set $\left(T'\cup T''\right)\setminus\{\tilde\gamma_s\}$ all tuples of
maximum weight are contained in $T''$. Note that from any pair of antichains
$(T', T'')\in {\cal A}_s$ containing tuples with weight less than~$s$ we can
obtain a regular from below pair of antichains by placing in the antichain $T'$
all tuples from $\left(T'\cup T''\right)\setminus\{\tilde\gamma_s\}$ which have 
the minimum weight. We call the pair of antichains obtained by this way from
the initial pair $(T', T'')$ the {\em correction from below} of $(T', T'')$.
In an analogous way, from any pair of antichains $(T', T'')\in {\cal A}_s$
we can obtain a regular from above pair of antichains by placing to the antichain $T''$
all tuples from $\left(T'\cup T''\right)\setminus\{\tilde\gamma_s\}$ which 
have the maximum weight. We call the pair of antichains obtained by this way
the {\em correction from above} of the initial pair $(T', T'')$. Note that
both the correction from below and the correction from above consist of the
same tuples as the initial pair of antichains.

\begin{lem}
For any $s>n/2$ in ${\cal A}_s$ there exists a pair of antichains which has
the maximum cardinality and consists of tuples with weight greater than or equal
to $\lfloor n/2\rfloor$.
\label{keylem1}
\end{lem}

{\bf Proof.}
Consider an arbitrary pair of antichains $(T', T'')$ from ${\cal A}_s$ which
has the maximum cardinality. Assume that the minimum weight of tuples from 
$T'\cup T''$ is equal to $r<\lfloor n/2\rfloor$. Let $(T'_0, T''_0)$ be
the correction from below of the pair $(T', T'')$. Since $T'_0\cup T''_0 = T'\cup T''$,
the pair of antichains $(T'_0, T''_0)$ has also the maximum cardinality
in ${\cal A}_s$, and the minimum weight of tuples from $(T'_0, T''_0)$
is also equal to $r$. Moreover, since $(T'_0, T''_0)$ is regular from below
and $r\neq s$, all tuples from $T'_0\cup T''_0$ whose weights are equal to~$r$
are contained in $T'_0$. Denote the set of all such tuples by~$V$. Denote
also by~$U$ the set of all tuples from $T''_0\cap B^n_{r+1}$. Furthermore,
denote by $V'$ the set of all tuples from $B^n_{r+1}$ which are comparable
with at least one tuple from~$V$, and by $U'$ the set of all tuples from $B^n_{r+2}$ 
which are comparable with at least one tuple from~$U$. Note that each tuple
from $B^n_r$ is comparable with $n-r$ tuples from $B^n_{r+1}$ and each tuple
from $B^n_{r+1}$ is comparable with $r+1$ tuples from $B^n_r$. From these
observations we conlude that $|V'|\ge\frac{n-r}{r+1}|V|$. In the analogous way 
we obtain that $|U'|\ge\frac{n-r-1}{r+2}|U|$. Note also that $V'\cap T'_0=\emptyset$
and $U'\cap T''_0=\emptyset$ because $T'_0$ and $T''_0$ are antichains.
First consider the case $r<n/2-1$, i.e. $r\le n/2-3/2$. In this case we
have $|V'|\ge\frac{n-r}{r+1}|V|>|V|$ and $|U'|\ge\frac{n-r-1}{r+2}|U|\ge |U|$.
Define $T'_1=(T'_0\setminus V)\cup V'$ and $T''_1=(T''_0\setminus U)\cup U'$. 
It is easy to note that $(T'_1, T''_1)\in {\cal A}_s$. Moreover, we have
$|T'_1|=|T'_0|+|V'|-|V|>|T'_0|$ and $|T''_1|=|T''_0|+|U'|-|U|\ge |T''_0|$.
Therefore, $|T'_1\cup T''_1|=|T'_1|+|T''_1|>|T'_0|+|T''_0|=|T'_0\cup T''_0|$,
which contradicts the fact that the pair of antichains $(T'_0, T''_0)$ has the maximum 
cardinality in ${\cal A}_s$. Thus the case $r<n/2-1$ is impossible.
Now consider the remaining case $r=n/2-1$. Note that in this case 
$n$ has to be even, i.e. $n=2k$, and $r=\lfloor n/2\rfloor -1=k-1$.
Denote by $V''$ the set $V'\cup U$. Note that $V''\cap T'_0=\emptyset$
because both the sets $V'$ and $U$ are not overlapped with $T'_0$.
Define $T'_2=(T'_0\setminus V)\cup V''$ and $T''_2=(T''_0\setminus U)\cup U'$.
It is easy to note that  $(T'_2, T''_2)\in {\cal A}_s$. Moreover,
we have $|V''|\ge |V'|\ge\frac{n-r}{r+1}|V|=\frac{k+1}{k}|V|$, i.e. 
$|V|\le \frac{k}{k+1}|V''|$. We have also $|U'|\ge\frac{n-r-1}{r+2}|U|=
\frac{k}{k+1}|U|$. Thus, taking into account that $|U|\le |V''|$, we obtain
$$
(|V''|-|V|)+(|U'|-|U|)\ge\frac{1}{k+1}|V''|-\frac{1}{k+1}|U|\ge 0.
$$
Therefore,
\begin{eqnarray*}
|T'_2\cup T''_2|&=&|T'_2|+ |T''_2|=|T'_0|+ |T''_0|+(|V''|-|V|)+(|U'|-|U|)\\
&\ge&|T'_0|+ |T''_0|=|T'_0\cup T''_0|,
\end{eqnarray*}
i.e. the cardinality of $(T'_2, T''_2)$ is not less than the cardinality of
$(T'_0, T''_0)$. Thus the pair of antichains $(T'_2, T''_2)$ has also the 
maximum cardinality in ${\cal A}_s$. Moreover, it is obvious that the 
antichains $T'_2, T''_2$ consist of tuples whose weights are not less than
$\lfloor n/2\rfloor$. So the lemma is proved. $\Box$

\begin{lem}
For any $s>n/2$ in ${\cal A}_s$ there exists a pair of antichains such that
this pair has the maximum cardinality and the weights of all tuples from these
antichains except the tuple $\tilde\gamma_s$ are not greater than $(n+3)/2$
and not less than $\lfloor n/2\rfloor$.
\label{keylem2}
\end{lem}

{\bf Proof.}
By Lemma~\ref{keylem1} there exists a pair of antichains $(T', T'')$  in ${\cal A}_s$
which has the maximum cardinality and consists of tuples with weight greater than or equal
to $\lfloor n/2\rfloor$. Assume that the maximum weight of tuples from these antichains 
except the tuple $\tilde\gamma_s$ is equal to $r> (n+3)/2$. Note that the inequality
$r> (n+3)/2$ obviously implies $r\ge n/2 +2$. For proving Lemma~\ref{keylem2} it is 
enough to show that in this case we can construct a pair of antichains from ${\cal A}_s$
such that this pair has the maximum cardinality in ${\cal A}_s$ and the weights of all 
tuples from these antichains except the tuple $\tilde\gamma_s$ are not less than $\lfloor n/2\rfloor$
and not greater than $r-1$. To this end, consider the correction from above of $(T', T'')$.
Denote this correction by $(T'_0, T''_0)$. Since $T'_0\cup T''_0=T'\cup T''$, the pair
of antichains $(T'_0, T''_0)$ has also the maximum cardinality in ${\cal A}_s$, and
the weights of all tuples from $T'_0\cup T''_0$ except the tuple $\tilde\gamma_s$ are not 
less than $\lfloor n/2\rfloor$ and not greater than~$r$. Moreover, all tuples from
$(T'_0\cup T''_0)\cap B^n_r$ except the tuple $\tilde\gamma_s$ are contained in $T''_0$.
Denote the set of all such tuples by~$U$. Denote also by~$V$ the set 
$(T'_0\cap B^n_{r-1})\setminus\{\tilde\gamma_s\}$. Furthermore, denote by ${\cal D}(U)$ 
(${\cal D}(V)$) the set $\{ {\cal D}(\tilde\alpha)\;:\;\tilde\alpha\in U\}$ 
($\{ {\cal D}(\tilde\alpha)\;:\;\tilde\alpha\in V\}$). It follows from Lemma~\ref{injectD}
that $|{\cal D}(U)|=|U|$ and $|{\cal D}(V)|=|V|$. Moreover, since $T'_0$ and $T''_0$ are
antichains, we have ${\cal D}(V)\cap T'_0=\emptyset$ and ${\cal D}(U)\cap T''_0=\emptyset$.
Define $T'_1=(T'_0\setminus V)\cup {\cal D}(V)$ and $T''_1=(T''_0\setminus U)\cup {\cal D}(U)$.
Using Lemma~\ref{nesravD1}, it is easy to check that $T'_1$ is an antichain containing
the tuple $\tilde\gamma_s$. Moreover, it is obvious that $T''_1$ is also an antichain.
By Lemma~\ref{nesravD1} any tuple $\tilde\alpha$ from $T''_1$ does not satisfy the relation
$\tilde\alpha\le\tilde\gamma_s$. Taking this observation into account, it is easy to see that $T'_1<T''_2$.
Thus $(T'_1, T''_1)\in {\cal A}_s$. It follows from $|{\cal D}(U)|=|U|$ and $|{\cal D}(V)|=|V|$
that $|T'_1|=|T'_0|$ and $|T''_1|=|T''_0|$, so the pair of antichains $(T'_1, T''_1)$  has also
the maximum cardinality in ${\cal A}_s$. To complete the proof, we note that the weight of any
tuple from $T'_1\cup T''_1$ except the tuple $\tilde\gamma_s$ are not less than $\lfloor n/2\rfloor$
and not greater than $r-1$. $\Box$

\begin{teor}
For $s>n/2$ the cardinality of any pair of antichains from ${\cal A}_s$ is not greater than 
$1+{ n+1 \choose \lfloor n/2\rfloor +1}-{s+1 \choose \lfloor n/2\rfloor +1}$.
\label{keyteor}
\end{teor}

{\bf Proof.}
First consider the case when $n$ is even, i.e. $n=2k$.
In this case, according to Lemma~\ref{keylem2}, there exists a pair of antichains
$(T', T'')$ in ${\cal A}_s$ such that this pair has the maximum cardinality in ${\cal A}_s$
and the weights of all tuples from $(T'\cup T'')\setminus\{\tilde\gamma_s\}$ are either $k$ 
or $k+1$. Therefore, for any tuple $\tilde\alpha$ from $T''$ the relation $\tilde\alpha\ge\tilde\gamma_s$
can not be valid because the weight of $\tilde\gamma_s$ is equal to $s\ge k+1$. So all
tuples from $T''$ are incomparable with $\tilde\gamma_s$. Since $T'$ is a antichain
containing $\tilde\gamma_s$, all tuples from $T''$ are also incomparable with $\tilde\gamma_s$.
Thus, all tuples from $T'\cup T''\setminus\{\tilde\gamma_s\}$ are incomparable with $\tilde\gamma_s$. 
It is obvious that $B^n_k$ ($B^n_{k+1}$) contains ${n \choose k}-{s \choose k}$
(${n \choose k+1}-{s \choose k+1}$) tuples incomparable with $\tilde\gamma_s$.
Hence
$$
|T'\cup T''\setminus\{\tilde\gamma_s\}|\le 
\left({n \choose k+1}-{s \choose k+1}\right)+\left({n \choose k}-{s \choose k}\right)={n+1 \choose k+1}-{s+1 \choose k+1}.
$$
Therefore, $|T'\cup T''|\le 1+{n+1 \choose k+1}-{s+1 \choose k+1}$.
Since the pair $(T', T'')$ has the maximum cardinality in ${\cal A}_s$,
we obtain that in this case the cardinality of any pair of antichains
from ${\cal A}_s$ is not greater than
$$
1+{n+1 \choose k+1}-{s+1 \choose k+1}=1+{n+1 \choose \lfloor n/2\rfloor+1}-{s+1 \choose \lfloor n/2\rfloor+1}.
$$

Now consider the case when $n$ is odd, i.e. $n=2k+1$. By Lemma~\ref{keylem2}, in this case
there exists a pair of antichains $(T', T'')$ in ${\cal A}_s$ such that this pair has the 
maximum cardinality in ${\cal A}_s$ and the weights of all tuples from $(T'\cup T'')\setminus\{\tilde\gamma_s\}$ 
can be equal to three posssible values: $k$, $k+1$, or $k+2$. Let $(T'_0, T''_0)$ be
the correction from above of $(T', T'')$. Since $T'_0\cup T''_0 = T'\cup T''$, the
pair $(T'_0, T''_0)$ has also the maximum cardinality in ${\cal A}_s$ and the weights of all 
tuples from $(T'_0\cup T''_0)\setminus\{\tilde\gamma_s\}$ can be equal to $k$, $k+1$, or $k+2$.
Moreover, all tuples from $((T'_0\cup T''_0)\setminus\{\tilde\gamma_s\})\cap B^n_{k+2}$ are
contained in $T''_0$. Denote the set of all such tuples by~$U$. Define 
${\cal D}(U)=\{ {\cal D}(\tilde\alpha)\;:\;\tilde\alpha\in U\}$, $V={\cal D}(U)\cap T'_0$, 
and ${\cal D}(V)=\{ {\cal D}(\tilde\alpha)\;:\;\tilde\alpha\in V\}$.
Since $T'_0, T''_0$ are antichains, we have ${\cal D}(U)\cap T''_0=\emptyset$ and
${\cal D}(V)\cap T'_0=\emptyset$. Denote by $T'_1$ the set $(T'_0\setminus V)\cup {\cal D}(V)$
and by $T''_1$ the set $(T''_0\setminus U)\cup {\cal D}(U)$. It is easy to note
that $T'_1\cap T''_1=\emptyset$ and the weight of any tuple from $T'_1\cup T''_1\setminus\{\tilde\gamma_s\}$
is either $k$ or $k+1$. Taking into account that the weight of $\tilde\gamma_s$ is equal to $s\ge k+1$,
we obtain that for any tuple $\tilde\alpha$ from $T'_1\cup T''_1\setminus\{\tilde\gamma_s\}$
the relation $\tilde\alpha\ge\tilde\gamma_s$ can not be valid. Moreover, the relation 
$\tilde\alpha\le\tilde\gamma_s$ can not be valid for any tuple $\tilde\alpha$ from $T''_1$ 
by Lemma~\ref{nesravD1} and can not be also valid for any tuple $\tilde\alpha$ from 
$T'_1\setminus\{\tilde\gamma_s\}$ by Lemma~\ref{nesravD2}. Thus, all tuples from 
$T'_1\cup T''_1\setminus\{\tilde\gamma_s\}$ are incomparable with $\tilde\gamma_s$.
Hence, by the same way as in the previous case of even~$n$, we obtain that 
$$
|T'_1\cup T''_1\setminus\{\tilde\gamma_s\}|\le {n+1 \choose k+1} - {s+1\choose k+1}.
$$
Therefore
$$
|T'_1\cup T''_1|\le 1+{n+1 \choose k+1}-{s+1 \choose k+1}=1+{n+1 \choose \lfloor n/2\rfloor +1}-
{s+1 \choose \lfloor n/2\rfloor +1}.
$$
Lemma~\ref{injectD} implies $|{\cal D}(U)|=|U|$ and $|{\cal D}(V)|=|V|$. Hence $|T'_1|=|T'_0|$ and 
$|T''_1|=|T''_0|$, so $|T'_0\cup T''_0|=|T'_1\cup T''_1|$. Therefore
$$
|T'_0\cup T''_0|\le 1+{n+1 \choose \lfloor n/2\rfloor +1} - {s+1 \choose \lfloor n/2\rfloor +1}.
$$
Thus, since the pair of antichains $(T'_0, T''_0)$ has the maximum cardinality in ${\cal A}_s$,
we obtain that in this case also the theorem is valid. $\Box$

\begin{sled}
Let $s>n/2$ and $(T', T'')$ be the pair of antichains such that $T'$ consists of the tuple $\tilde\gamma_s$
and all tuples from $B^n_{\lfloor n/2\rfloor}$ which are incomparable with $\tilde\gamma_s$
and $T''$ consists of all tuples from $B^n_{\lfloor n/2\rfloor+1}$ which are incomparable 
with $\tilde\gamma_s$. Then $(T', T'')$ has the maximum cardinality in ${\cal A}_s$.
\label{keysled}
\end{sled}

Denote by ${\cal A}'_t$ the set of all pairs $(T', T'')$ of antichains in $B^n$ such that $T'<T''$ 
and the weights of all tuples from $T'$ and $T''$ are no more than~$t$.

\begin{teor}
For $t\le \lfloor n/2\rfloor +1$ the cardinality of any pair of antichains from ${\cal A}'_t$
is not greater than ${n+1 \choose t}$.
\label{scdteor}
\end{teor}

{\bf Proof.}
Let $(T', T'')$ be an arbitrary pair of antichains from ${\cal A}'_t$ and $q$
be the cardinality of $(T', T'')$. A chain of tuples in $B^n$
is called {\em maximal} if it consists of $n+1$ tuples. For any tuple in $B^n$
we consider the number of different maximal chains containing this tuple.
We will call this number the {\em rank} of the tuple. It is easy to check
that the rank of a tuple is equal to $k!(n-k)!$ where $k$ is the weight of
the tuple. Note that in $B^n$ there exist $n!$ different maximal chains and
each of these chains contains no more than one tuple from $T'$ and no more 
than one tuple from $T''$. So the total sum of ranks of all tuples from $T'\cup T''$
is no more than $2(n!)$. Consider the sequence of all tuples in $B^n$ whose weights are
no more than~$t$ such that in this sequence tuples are sorted in the
non-decreasing  order of their ranks. Denote this sequence by~$H$. It is obvious that
the sum of ranks of all tuples from $T'\cup T''$ is not less than the sum
of ranks of the first~$q$ tuples in~$H$. Thus the sum of ranks of the first~$q$ tuples 
in~$H$ is also not greater than $2(n!)$. 

First consider the case $t\le\frac{n+1}{2}$. Note that the value $k!(n-k)!$ is not 
increasing for $0\le k\le (n+1)/2$, so in this case we can assume that in~$H$ the first 
${n \choose t}$ tuples are tuples from $B^n_t$ and the following  ${n \choose t-1}$ tuples 
are tuples from $B^n_{t-1}$. Thus in~$H$ the sum of ranks of the first 
${n \choose t}+{n \choose t-1}={n+1 \choose t}$ tuples is equal to
$$
{n \choose t}\cdot t!(n-t)!+{n \choose t-1} \cdot (t-1)!(n-t+1)!=2(n!).
$$
Therefore, $q$ can not be greater than ${n+1 \choose t}$.
Now consider the case $t=n/2+1$ which is possible only for even~$n$.
Note that in this case ${n \choose n/2}$ tuples from $B^n_{n/2}$ have
the minimal rank $(n/2)!(n/2)!$ in~$H$ while all the other tuples in~$H$
have ranks not less than $(n/2-1)!(n/2+1)!$. So we can assume that in~$H$ the first 
${n \choose n/2}$ tuples are tuples from $B^n_{n/2}$ and the following  
${n \choose n/2+1}$ tuples are tuples from $B^n_{n/2+1}$. Thus in~$H$ 
the sum of ranks of first ${n \choose n/2}+{n \choose n/2+1}={n+1 \choose n/2+1}$
tuples is equal to
$$
{n \choose n/2}\cdot (n/2)!(n/2)!+{n \choose n/2+1}\cdot (n/2-1)!(n/2+1)!=2(n!).
$$
Therefore, in this case also $q$ can not be greater than ${n+1 \choose n/2+1}=
{n+1 \choose t}$. $\Box$

\begin{sled}
Let $s\le \lfloor n/2\rfloor +1$ and $(T', T'')$ be the pair of antichains such that 
$T'$ consists of all tuples from $B^n_{t-1}$ and $T''$ consists of all tuples from 
$B^n_{t}$. Then $(T', T'')$ has the maximum cardinality in ${\cal A}'_t$.
\label{scdsled}
\end{sled}

\section{The MBnB complexity bounds}
\label{sec:compl}

Now we obtain upper bounds for the MBnB complexity of the problem~(\ref{eq:ssum}) from
the statements, proved in Section~\ref{sec:pre}, and Theorems \ref{keyteor} and~\ref{scdteor}.
Denote by $T_0$ ($T_1$) the $0$-antichain ($1$-antichain) for the problem~(\ref{eq:ssum}).
Define the values $t$ and $s$ in the following way:
\begin{equation}
\label{eq:ts}
t = \min\left\{k \in N: \sum_{i=n - k + 1}^n w_i > C\right\},\,\,\, s = t - 1.
\end{equation}

We prove the following 

\begin{utv}
\label{utv:t0t1}
The weight of any tuple from $T_0$ and $T_1$ is no greater than~$t$, and $\tilde\gamma_s \in T_1$.
\end{utv}
{\bf Proof.}
Consider a $1$-tuple  $\tilde\alpha=(\alpha_1,\ldots,\alpha_n)$ from~$T_1$. By Proposition~\ref{utv:c1}
we have $\sum_{i=1}^n \alpha_i w_i \leq C$. Since $w_1 \geq \dots \geq w_n$, the inequality 
$\sum_{i=n-\|\tilde\alpha\|+1}^n w_i\leq \sum_{i=1}^n \alpha_i w_i$ is valid, so 
$\sum_{i=n-\|\tilde\alpha\|+1}^n w_i\leq C$. Therefore $\|\tilde\alpha\|<t$.
Now consider a $0$-tuple $\tilde\beta=(\beta_1,\ldots,\beta_n)$ from~$T_0$.
Let $j = \max \{i \in N : \beta_i = 1\}$. By the definition of a leaf $0$-tuple 
we have $\sum_{i = 1}^{j-1} \beta_i w_i\leq C$. The inequalities $w_1 \geq \dots \geq w_n$
imply that $\sum_{i=n-\|\tilde\beta\|+2}^n w_i\leq \sum_{i = 1}^{j-1} \beta_i w_i$.
Hence $\sum_{i=n-\|\tilde\beta\|+2}^n w_i\leq C$. Therefore, $\|\tilde\beta\|-1<t$,
so $\|\tilde\beta\|\leq t$.

Now we prove that $\tilde\gamma_s\in T_1$. Consider the subproblem~$P$ corresponding to the map $(I,\theta)$
such that $I = \{1, \dots, n - s-1\}$ and $\theta (i) = 0, i \in I$. For this subproblem we have
$$
\sum_{i\in I} (1 - \theta_i) w_i = \sum_{i=1}^{n - s - 1} w_i = W - \sum_{i=n - t + 1}^n w_i < W - C.
$$
Thus the subproblem~$P$ does not satisfy the C1-condition. It is obvious that $P$ does not satisfy also 
the C0-condition. We conclude from these observations that $P$ is contained in the MBnB-tree but is not
a leaf of this tree. Now consider the subproblem~$P'$ corresponding to the map $(I',\theta')$ such that
$I' = \{1, \dots, n - s \}$ and $\theta' (i) = 0, i \in I'$. For this subproblem we have
$$
\sum_{i\in I'} (1 - \theta'_i) w_i = \sum_{i=1}^{n - s} w_i = W - \sum_{i=n - s +1}^n w_i \geq W - C.
$$
Thus the subproblem~$P'$ satisfies the C1-condition. Moreover, $P'$ is obviously contained 
in the decomposition of the subproblem~$P$. Therefore, $P'$ is a leaf of the MBnB-tree satisfying 
the C1-condition. Note that $\tilde\gamma_s$ is the $1$-complement for the map $(I',\theta')$
corresponding for~$P'$, so $\tilde\gamma_s$ is a leaf $1$-tuple.$\Box$

From Propositions \ref{utv:t0t1} and~\ref{utv:01chains} we obtain that the pair of antichains $(T_1, T_0)$ is
contained in the set ${\cal A}_s$ if $t>\lfloor n/2\rfloor +1$ or in the set ${\cal A}'_t$ if $t\leq \lfloor n/2\rfloor +1$.
So Theorems \ref{keyteor},~\ref{scdteor} imply the following bounds for the MBnB complexity of the subset sum problem.

\begin{teor}
\label{mainres}
The MBnB complexity~$S$ of the problem (\ref{eq:ssum}) satisfies the following upper bounds:
\[
\begin{array}{lcl}
\cal S & \leq & 2 {n+1 \choose t} - 1, \mbox{ if } t\le \lfloor n/2\rfloor +1,\\
\\
\cal S & \leq & 2 \left({ n+1 \choose \lfloor n/2\rfloor +1}-{t \choose \lfloor n/2\rfloor +1}\right) + 1, \mbox{ if } t> 
\lfloor n/2\rfloor + 1,
\end{array}
\]
where $t$ is defined in~(\ref{eq:ts}).
\end{teor}

\section{Comparison of bounds}
\label{sec:exp}

In this section we compare the known complexity bounds with
the complexity bound proposed in this paper:

\begin{center}
{\renewcommand{\arraystretch}{2}
\begin{tabular}{|c|c|c|}
\hline
Designation & Formula & Source\\
\hline
B1 & $2 {n + 1 \choose \lfloor n/2 \rfloor + 1} - 1$ &  
Girshukhin~\cite{Grish}\\
\hline
B2 & $2 { n + 1 - t' + t \choose t} - 1$ & 
Kolpakov and Posypkin~\cite{Kolpak2}\\
\hline
B3 &$\begin{array}{l}
2 {n+1 \choose t} - 1 \mbox{ if } t\le \lfloor n/2\rfloor +1;\\
\\
2 \left({ n+1 \choose \lfloor n/2\rfloor +1}-{t \choose \lfloor n/2\rfloor +1}\right) + 1\\
 \mbox{ if } t> \lfloor n/2\rfloor + 1
\end{array}$ & this paper\\
\hline
\end{tabular}
}
\end{center}
Parameters $t$ and $t'$ are computed as follows:
\[
t = \min \left\{k\in N : \sum_{i = n-k+1}^n w_i > C\right\},
t' = \min \left\{k\in N: \sum_{i = 1}^k w_i > C\right\}.
\]

It is obvious that bound B3 is better than B1.
As for comparison of B2 and B3, in the case of 
$t\le \lfloor n/2\rfloor + 1$ bound B3 coinsides
with B2 for $t=t'$ and is better than B2 for $t>t'$
(note that $t\ge t'$). In the case of $t>\lfloor n/2\rfloor + 1$
bound B2 may be better under some conditions. However,
in this case bound B3 is better than B2 for $t'\le\lfloor n/2\rfloor + 1$
since this condition implies that
$$
\begin{array}{c}
\displaystyle
{n+1+t-t'\choose t}\ge {\lceil n/2\rceil + t\choose t} =
{\lceil n/2\rceil + t\choose \lceil n/2\rceil}>
{\lceil n/2\rceil + \lfloor n/2\rfloor + 1\choose \lceil n/2\rceil}=\\
\displaystyle
{n + 1\choose \lceil n/2\rceil}={n + 1\choose \lfloor n/2\rfloor +1}>
{n + 1\choose \lfloor n/2\rfloor +1}-{t\choose \lfloor n/2\rfloor +1}
\end{array}
$$

We also performed experimental comparison of bounds B1, B2 and B3.
For our experiments $1000$ subset sum instances were generated. Each instance had $15$ variables. 
Coefficients $w_i$ were uniformly distributed pseudo-random numbers in $[1, 100]$, 
$C$ was choosen randomly in $[1, \sum_{i=1}^n w_i]$. 
All instances were solved with MBnB algorithm. The average complexity of MBnB was $2114.02$. 
Table \ref{tab:compare} compares various bounds using the following indicators:

{\bf Average value}: the value of the bound averaged over all instances;

{\bf Min (Max) ratio}: the minimum (maximum) value of the scaled accuracy of the bound with 
respect to the actual number of steps performed by MBnB, computed as follows: $r = \frac{S' - S}{S}$, 
where $S$ and $S'$ are the actual complexity and the bound respectively;

{\bf Best bound}: the number of times when the bound gives the least value from all 3 bounds;

{\bf Precise bound}: the number of times when the bound was precise, i.e. equal to the actual complexity.

\begin{table}[h]
\caption{Comparison of various bounds}
\begin{center}
\begin{tabular}{|l|c|c|c|}
\hline
Indicator & B1 & B2 & B3\\
\hline
Average value & 25739 & 859985.552 & 20257.82\\
\hline
Min ratio & 0.908 & 0 & 0\\
\hline
Max ratio & 1224.667 & 7578.711 & 761.619\\
\hline
Best bound & 64 & 72 & 931\\
\hline
Precise bound & 0 & 15 & 3\\
\hline
\end{tabular}
\end{center}
\label{tab:compare}
\end{table}

The performed comparison shows that B3 bound outperforms bounds B1 and B2 in terms of average value and maximal relative accuracy. 
Bound B1 is data independent and thus the probability that it equals to the actual complexity is very low. 
Bound B2 gives the precise bound more often than B3. 
We should also take into account that B2 is a generic bound, suitable for a broad class of Branch-and-Bound methods, 
while B1 and B3 only work for MBnB. Experiments show that all three bounds make sense and can be applied in combination.

\section{Conclusion}
\label{sec:conc}

In this paper we obtained upper bounds on the complexity of solving the subset sum problem
by the Branch-and-Bound method where all subproblems are partitioned along the free variable
with the maximum weight. These bounds can be easily computed from the input data of the problem.
So these bounds allow preliminarily estimates of the number of operations required for
solving the problem. Such bounds can be used in planning of distributed computations,
for which one needs to estimate computational resources required for solving the problem.

For the obtained bounds a natural question arises: whether these bounds are tight? 
We can show that the obtained bounds are tight for $t\leq \lfloor n/2\rfloor +2$.
For $t\leq \lfloor n/2\rfloor +1$ these bounds are reached for the subset sum problem
with the parameters $w_1=w_2=\ldots =w_n=2$ and $C=2t-1$ (see Kolpakov and Posypkin~\cite{Kolpak1}).
For $t= \lfloor n/2\rfloor +2$ these bounds are reached for the subset sum problem
with the parameters $w_1=w_2=\ldots =w_{n-k-1}=3k$, $w_{n-k}=w_{n-k+1}=\ldots =w_n=3k-2$,
and $C=3k^2+k-1$ where $k=\lfloor n/2\rfloor$. On the other hand, we can show that
for $n=7$ and $t=6$ the MBnB complexity of the subset sum problem is not greater
than 53 while the complexity upper bound derived in this case from Theorem~\ref{mainres}
is 56. Thus, just for $t= \lfloor n/2\rfloor +3$ the obtained upper bounds are not
exact, so one of the directions for further research is to improve the obtained bounds
for the case $t>\lfloor n/2\rfloor +2$. We also intend to improve these bounds for the 
boolean knapsack problem and obtain lower bounds for the considered problem.


\bibliographystyle{ws-ijfcs}
\bibliography{kolposArhiv}

\begin{thebibliography}{10}

\bibitem{Aardal}
K.~Aardal and A.~Lenstra, Hard equality constrained integer knapsacks, {\em
  Mathematics of Operations Research} {\bf 29}(3)  (2004)  724--738.

\bibitem{Chvatal}
V.~Chvatal, Hard knapsack problems, {\em Operations Research} {\bf 28}(6)
  (1980)  1402--1411.

\bibitem{Finkel}
Y.~Finkelshtein, {\em Approximate methods and applied problems in discrete
  programming (in Russian)} (Nauka, 1976).

\bibitem{Greenberg}
H.~Greenberg and R.~Hegerich, A branch and bound algorithm for the knapsack
  problem, {\em Management Science} {\bf 16}(5)  (1970)  327--332.

\bibitem{Grish}
V.~Grishuhin, The efficiency of branch-and-bound method in boolean programming
  (in russian), {\em Research on discrete optimization\/},  ed. A.~Fridman
  (Nauka, 1976), pp. 203--230.

\bibitem{Jeroslow}
R.~Jeroslow, Trivial integer programs unsolvable by branch-and-bound, {\em
  Mathematical Programming} {\bf 6}  (1974)  105--109.

\bibitem{Kolpak1}
R.~Kolpakov and M.~Posypkin, An asymptotic bound on the complexity of the
  branch-and-bound method with branching by the fractional variable in the
  knapsack problem, {\em Diskret. Anal. Issled. Oper.} {\bf 15}(1)  (2008)
  58--81.

\bibitem{Kolpak2}
R.~Kolpakov and M.~Posypkin, Upper and lower bounds for the complexity of the
  branch and bound method for the knapsack problem, {\em Discrete Mathematics
  and Applications} {\bf 20}(1)  (2010)  1569--3929.

\bibitem{Krishna}
B.~Krishnamoorthy, Bounds on the size of branch-and-bound proofs for integer
  knapsacks, {\em OR Letters} {\bf 36}(1)  (2008)  19--25.

\bibitem{Martello}
S.~Martello and P.~Toth, {\em Knapsack Problems: Algorithms and Computer
  Implementation} (John Wiley and Sons, 1990).

\bibitem{Kellerer}
H.~K.~U. Pfershy and D.~Pisinger, {\em Knapsack Problems} (Springer, 2004).

\end{thebibliography}

\end{document}